\documentclass[twocolumn,showpacs,nofootinbib]{revtex4-1}
\usepackage{epsfig,amssymb,amsmath,latexsym}
\usepackage{pifont}
\usepackage{ulem}
\usepackage{hyperref}
\usepackage{subfig}
\usepackage{multirow, makecell}

\usepackage{graphicx}
\usepackage{dcolumn}
\usepackage{bm}
\usepackage{color}

\hypersetup{
    colorlinks=true,
    linkcolor=red,
    citecolor=blue,
    urlcolor=black
}

\definecolor{nicegreen}{rgb}{0.1,0.5,0.1}
\definecolor{nicered}{rgb}{0.7,0.1,0.1}
\definecolor{amethyst}{rgb}{0.6, 0.4, 0.8}

\newcommand{\edits}[1]{{#1}}

\def\fun#1#2{\lower3.6pt\vbox{\baselineskip0pt\lineskip.9pt
  \ialign{$\mathsurround=0pt#1\hfil##\hfil$\crcr#2\crcr\sim\crcr}}}
\def\simgt{\mathrel{\lower0.6ex\hbox{$\buildrel {\textstyle >}
 \over {\scriptstyle \sim}$}}}
\def\simlt{\mathrel{\lower0.6ex\hbox{$\buildrel {\textstyle <}
 \over {\scriptstyle \sim}$}}}

\def\bea{\begin{eqnarray}}
\def\eea{\end{eqnarray}}
\def\be{\begin{equation}}
\def\ee{\end{equation}}

\input epsf

\def\ba{\begin{eqnarray}}
\def\ea{\end{eqnarray}}

\def\nn{\nonumber}

\def\FIX{\textsf{Fixed-sky}}
\def\FULL{\textsf{Full-sky}}
\def\HYB{\textsf{Informed-sky}}

\makeatletter
\long\def\@makecaption#1#2{%
  \par
  \vskip\abovecaptionskip
  \begingroup
    \small\rmfamily
    \flushing
    \let\footnote\@footnotemark@gobble
    #1\@caption@fignum@sep#2\par
  \endgroup
  \vskip\belowcaptionskip
}
\makeatother

\begin{document}

\preprint{}

\title{Impact of sky localization uncertainty on ringdown inference}

 \author{Kallol Dey${}^{1,2,3}$}
 \author{Enrico Barausse${}^{2,3}$}
 \author{Marco Crisostomi${}^{4,5}$}
 \author{Roberto Trotta${}^{2,6,7}$}

\affiliation{${}^{1}$ eXtreme Gravity Institute, Department of Physics, Montana State University, Bozeman, Montana 59717, USA}
\affiliation{${}^{2}$ SISSA, Via Bonomea 265, 34136 Trieste, Italy \& INFN Sezione di Trieste}
\affiliation{${}^{3}$ IFPU - Institute for Fundamental Physics of the Universe, Via Beirut 2, 34014 Trieste, Italy}
\affiliation{${}^{4}$ Dipartimento di Fisica, Universit\`a di Pisa, Largo B. Pontecorvo 3, 56127 Pisa, Italy}
\affiliation{${}^{5}$ INFN Sezione di Pisa, Largo B. Pontecorvo 3, 56127 Pisa, Italy}
\affiliation{${}^{6}$ ICSC -- Centro Nazionale di Ricerca in High Performance Computing, Big Data e Quantum Computing, Via Magnanelli 2, Bologna, Italy}
\affiliation{${}^{7}$ Physics Department, Blackett Lab, Imperial College London, Prince Consort Road, London SW7 2AZ, UK}

\date{\small \today}

\begin{abstract}
As gravitational-wave ringdown signals grow louder, quasinormal-mode inference depends increasingly on the treatment of extrinsic parameters. Standard analyses fix sky localization---and sometimes also polarization and inclination---to point estimates from a prior inspiral-merger-ringdown analysis, artificially breaking degeneracies and underestimating the true uncertainty of mode-amplitude values. We test two alternatives: uninformative priors on the extrinsic parameters, sampled jointly with the remnant mass, spin, mode amplitudes, and phases; and informed priors on sky position from the full signal posterior. The former yields wider marginal constraints on amplitude posteriors, and both avoid potential bias introduced by fixing the sky localization. In contrast, mode amplitude ratios remain consistent across approaches, making them a robust observable for Kerr spectroscopy. Our publicly available pipeline enables fast ringdown analyses capable of sampling {\it all} parameters, requiring tens of minutes on a laptop for a full inference. Applied to GW250114 and GW190521, our methods confirm the robust detection of the $(2,2,1)$ overtone in GW250114, and, for GW190521, find only mild evidence for the $(3,3,0)$ mode.
\end{abstract}


\maketitle

\section{Introduction}\label{sec:intro}

A perturbed black hole rings down by emitting quasinormal modes (QNMs), with complex frequencies set by its mass and spin in general relativity (GR). Measuring QNMs and testing consistency with the Kerr metric is the program of black hole spectroscopy~\cite{Dreyer:2003bv,Berti:2005ys,Berti:2009kk,Berti:2016lat,Bhagwat:2021kwv,Berti:2025hly}.

GW150914 inaugurated observational ringdown studies. Building on the expectation of Ref.~\cite{Giesler:2019uxc} that overtones can dominate the early post-merger, Ref.~\cite{Isi:2019aib} first claimed identification of the $(\ell,m,n)=(2,2,1)$ overtone, sparking debate on its robustness~\cite{Cotesta:2022pci,Isi:2022mhy}. Overtone evidence depends on the ringdown start time/validity regime, noise modeling, and statistical methodology~\cite{Cotesta:2022pci,Finch:2022ynt,Crisostomi:2023tle,Wang:2023ljx}. Ref.~\cite{Correia:2023bfn} showed that marginalizing over coalescence time and sky location---fixed in earlier analyses---reduces the overtone Bayes factor from $\sim 100$ to $O(1)$, revealing that the $(2,2,1)$ evidence in GW150914 was largely an artifact of fixing extrinsic parameters.

Higher-mode evidence has also been claimed: Ref.~\cite{Capano:2021etf} reported the $(3,3,0)$ mode in GW190521, and GWTC-3 found hints of the $(3,3,0)$ and $(2,1,0)$ modes when using the entire post-merger~\cite{Gennari:2023gmx,LIGOScientific:2021sio}. The GWTC-4 analysis is consistent with GR and highlights promising prospects for higher-mode detections as sensitivity improves~\cite{LIGOScientific:2026wpt}.
The loud GW250114 signal further transformed the landscape: it requires at least two QNMs, with  $(2,2,0)$ and $(2,2,1)$ frequencies matching Kerr at  tens-of-percent level~\cite{LIGOScientific:2025rid}, and shows possible evidence for the $(4,4,0)$ mode~\cite{LIGOScientific:2025wao} and   nonlinear QNMs~\cite{Wang:2026rev}. 

For such loud signals, the treatment of extrinsic parameters deserves scrutiny. The amplitudes $A_{\ell m n}$ [cf.~\eqref{template}] \edits{are degenerate with} inclination $\iota$, sky position $(\alpha,\delta)$, and polarization $\psi$ through antenna patterns and spin-weighted spherical harmonics. Fixing $(\alpha,\delta)$ (and sometimes $\psi$) to inspiral-merger-ringdown (IMR) point estimates~\cite{LIGOScientific:2025rid, LIGOScientific:2025wao} removes degeneracies and yields misleadingly narrow amplitude and phase posteriors. Pre-GWTC-3 analyses, in particular of GW150914~\cite{Isi:2019aib,Cotesta:2022pci,Isi:2022mhy}, fixed all extrinsic parameters, including $\iota$, removing the face-on/off degeneracy\footnote{\edits{Ref.~\cite{Carullo:2019flw} does not fix the extrinsic parameters, however it does not consider the sky-dependent inter-detector time delay in the likelihood.}}.

Wavelet-based frequency-domain approaches~\cite{Finch:2021qph,Finch:2022ynt} marginalize over the extrinsic parameters, but are computationally expensive. 
The $\mathcal{F}$-statistic~\cite{Wang:2024jlz} maximizes analytically over amplitudes and phases, but ignores  extrinsic-parameter uncertainties.
\edits{Inspiral--merger information can also enter ringdown inference through IMR templates~\cite{CalderonBustillo:2020rmh, Chandra:2025ipu} or IMR-informed priors~\cite{Wang:2026rev, Guo:2026oea}, and Ref.~\cite{Correia:2023bfn} modeled the pre-merger data with an IMR waveform to marginalize over the sky position.}

In this study, we investigate the impact of the sky-position treatment on ringdown inference, comparing three approaches:
\begin{itemize}
\item \FIX: the current standard analysis \cite{LIGOScientific:2021sio, LIGOScientific:2026wpt}, with sky position fixed at IMR point estimates; 
\item \FULL: an analysis with uninformative priors on all extrinsic parameters, estimated jointly with  remnant mass, spin, mode amplitudes, and phases;
\item \HYB: an inspiral-merger-informed analysis using the IMR sky posterior as prior, rather than a point estimate.
\end{itemize}

We validate the methods on simulations and apply them to GW250114 and GW190521 events. We find that fixing the sky position underestimates the uncertainty on  mode amplitudes, while amplitude \textit{ratios} are less sensitive to the sky-position prior\footnote{In the noiseless limit, extrinsic-parameter factors cancel exactly in same-$(\ell,m)$ \edits{strain} amplitude ratios; however, in cross-$(\ell,m)$ ones they do not, since the spin-weighted spherical harmonics differ. \label{foot1}}. Our results highlight the importance of marginalizing over extrinsic parameters as ringdown spectroscopy moves toward precision tests.

\edits{The paper is organized as follows: in Section \ref{sec:methodology}, we present our new methods alongside the standard one; in Section \ref{sec:results}, we present the results of their application to GW250114 and GW190521; and in Section \ref{sec:conclusions}, we draw our conclusions. Appendix \ref{app:sims} presents the validation of our pipeline using pure-ringdown and full IMR simulations.}

\section{Methodology}\label{sec:methodology} 

\begin{figure*}
    \centering
    \includegraphics[width=\linewidth]{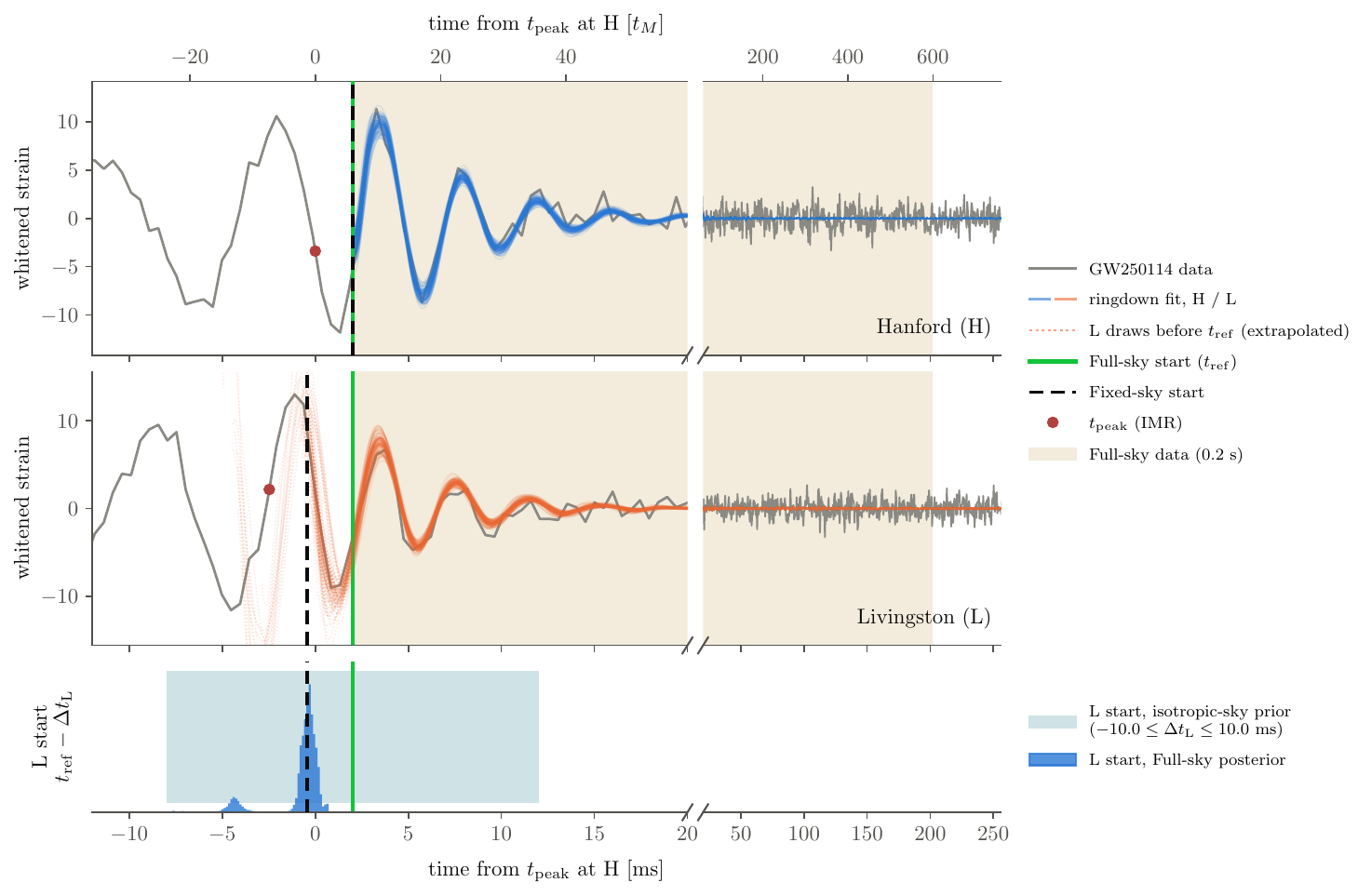}
    \caption{\footnotesize
    \edits{Data from LIGO Hanford (top) and LIGO Livingston (middle), together with the GW250114 ringdown signal reconstruction. The panels show whitened data as a function of time and signal reconstructions using 100 \FULL\ posterior draws of the $(2,2,0)+(2,2,1)$ fit. The shaded regions indicate the data used in the \FULL\ analysis, which are truncated (solid green lines) in both detectors at the same reference time $t_{\rm ref} = t_{\rm peak} + 6\,t_M$, defined at Hanford. The dashed black lines indicate the \FIX\ start times, which differ between detectors. The Livingston fit extrapolated before $t_{\rm ref}$ is shown as a dotted orange line. The bottom panel shows the posterior distribution of the Livingston start time, computed as $t_{\rm ref} - \Delta t_{\rm L}(\alpha,\delta)$, together with the range (shaded region) allowed by the isotropic prior on the sky position parameters.}
    }
    \label{fig:full-sky-demo}
\end{figure*}

We model the ringdown as a  superposition of QNMs indexed by $(\ell,m,n)$\footnote{This standard template assumes equatorial symmetry and neglects mode-mixing.}:
\ba
h_+(t) &+& i\, h_\times(t) = e^{-t/\tau_{\ell m n}} \Big[ {}_{-2}Y_{\ell m}(\iota)\, \mathcal{A}_{\ell m n}\, e^{i 2\pi f_{\ell m n} t} \nn\\
&+& (-1)^\ell\, {}_{-2}Y_{\ell,-m}(\iota)\, \mathcal{A}^*_{\ell m n}\, e^{-i 2\pi f_{\ell m n} t} \Big]\,,\label{template}
\ea
where $\mathcal{A}_{\ell m n} = A_{\ell m n}\, e^{i\phi_{\ell m n}}$ is a complex amplitude, ${}_{-2}Y_{\ell m}$ are spin-weighted spherical harmonics, and $f_{\ell m n}$, $\tau_{\ell m n}$ depend only on the remnant mass $M$ and spin $\chi$~\cite{Berti:2009kk}. The strain at detector $I$ is modeled as
\be\label{eq:hdet}
h_I(t) = F_{+,I}\, h_+(t + \Delta t_I) + F_{\times,I}\, h_\times(t + \Delta t_I)\,,
\ee
where $F_{+,I}$, $F_{\times,I}$ are antenna patterns depending on $(\alpha,\delta,\psi)$, and $\Delta t_I$ is the arrival-time delay relative to the last-observing detector. For instance, if Hanford (H) last sees the merger, 
$\Delta t_{\rm H}=0$
and for Livingston (L)
\be\label{eq:delay}
\Delta t_{\rm L} = (\mathbf{r}_{\rm H} - \mathbf{r}_{\rm L})\cdot \hat{\mathbf{n}}(\alpha,\delta)/c\,,
\ee
with $\mathbf{r}_{\rm H,L}$ the detector positions and $\hat{\mathbf{n}}(\alpha,\delta)$ the propagation direction. \edits{Therefore, the Livingston template is evaluated $\Delta t_L$ later in the waveform than the Hanford template. For a visual representation, relative to GW250114, see Fig.~\ref{fig:full-sky-demo}.}

The log-likelihood is \edits{$\ln\mathcal{L} = -\frac{1}{2}\sum_I ( d_I - h_I | d_I - h_I)_I - \frac{1}{2}\sum_I \ln |2\pi \mathbf{C}_I| $}, where $(\cdot|\cdot)_I$ is the noise-weighted inner product for detector $I$, \edits{$\mathbf{C}_I$ is the covariance matrix} and the sum runs over H and L. The inner product is computed in the time domain: the data must be truncated to exclude pre-ringdown signal, and a frequency-domain approach would introduce spectral leakage/contamination from windowing~\cite{Carullo:2019flw, Isi:2021iql}. 

In the standard \FIX\ approach, $(\alpha,\sin\delta)$ are fixed at point estimates and the remaining parameters are sampled. In our \FULL\ analysis, all parameters are sampled, so the antenna patterns and H--L time delay vary across parameter space, coupling extrinsic and intrinsic parameters. \edits{This comes at a cost: when the sky position is free, a {\it common} reference time $t_{\rm ref}$ must be set and used to truncate the data at {\it all} detectors. To avoid contamination from the merger part of the signal, we set it based on the strain peak at the last-observing detector. This implies that \FULL\ loses $\Delta t_I$ of data before $t_{\rm ref}$ in the first-observing detector respect to the \FIX\ case (see Fig.~\ref{fig:full-sky-demo}). 
Indeed, when using \FIX, no ringdown data is lost because the time-of-arrival delay between detectors is known (as function of the sky position) and, once an optimal start time is chosen for one detector, the corresponding start time in the other detector is automatically determined and used to truncate its data accordingly.}
The \HYB\ analysis follows the same likelihood as the \FULL\ one, but uses the IMR posterior for $(\alpha, \sin\delta)$ as prior for the inference, preserving the correlations and multimodality of the IMR localization. This is statistically correct only if the IMR sky posterior is dominated by the inspiral-merger; otherwise, it would double-count the data. Therefore, in the following, we apply \HYB\ only to GW250114 and not to GW190521\footnote{GW190521 has only a few cycles in band, so the inspiral carries no significant information on sky position. \label{foot}}.
As customary in the \FIX\ analysis, we also consider multiple start times to account for the unknown ringdown onset, but in Figures~\ref{fig:GW250114_comp}--\ref{fig:GW190521_comp} we report results for one value.

\edits{We assume stationary Gaussian noise whose autocorrelation function for each detector is estimated in the time domain from 64 seconds of off-source data on each side of the event and averaged to the two estimates. From it, we build a (non-circulant) Toeplitz covariance matrix $\mathbf{C}_I$ and Cholesky-decompose $\mathbf{C}_I = \mathbf{L}_I \mathbf{L}_I^T$ to evaluate the inner product via
back-substitution.}

We adopt uniform priors: the remnant spin $\chi \in [0,0.99]$, $\phi_{\ell m n}\in [0,2\pi]$, $\cos \iota \in [-1,1]$, $\alpha \in [0,2\pi]$, $\sin\delta \in [-1,1]$, $\psi \in [0,\pi]$, \edits{$A_{2 2 0}$ and $A_{2 2 1} \in [0,4]\cdot 10^{-20}$, $A_{3 3 0} \in [0,0.6]\cdot 10^{-20}$}, and an event-dependent range for the mass $M$. We analyze $0.2 \, s$ of data sampled at $4096$ Hz. Inference uses nested sampling~\cite{Skilling:2006} via \texttt{jaxns}~\cite{Albert:2020jaxns} with \edits{$10^5$ live points}\footnote{\edits{With $10^4$ live points, the posteriors on the remnant mass, spin, mode amplitudes and their ratios, as well as the Bayes factors, agree with those using $10^5$ live points. The latter only feature a reduced sampling noise in the 2D contours. We therefore use $10^4$ live points for \HYB\ and for Figs.~\ref{fig:GW250114_violin},~\ref{fig:GW190521_violin} for computational efficiency.}} and tolerance $\Delta\ln\mathcal{Z} < 0.01$ \footnote{We  verified that setting $\Delta \ln \mathcal{Z}=10^{-4}$ has no appreciable impact on the inferred posteriors and Bayes factors.}. 

\section{Results}\label{sec:results}

We apply all  approaches to GW250114 (\HYB\ is applicable here because the IMR sky localization is set mainly by the pre-merger signal). The $(4,4,0)$ and quadratic modes remain debated, but  $(2,2,0)$  and  $(2,2,1)$  are strongly supported by previous analyses~\cite{LIGOScientific:2025rid, LIGOScientific:2025wao}; thus we use these two modes only. Hanford observes the signal after Livingston, so we use the Hanford timeseries to set the common reference GPS time  $t_{\rm ref}=1420878141.221026$, corresponding to $t_{\rm peak}+6 t_M$ at Hanford \cite{LIGOScientific:2025rid} (where $t_M = 68.409 \ M_\odot = 0.337$ ms, \edits{using $t_M =GM/c^3$}). For the \FIX\ analysis, we fix the sky location at $(\alpha, \delta)=(2.33, 0.19)$~\cite{LIGOScientific:2025rid}, set the same GPS time $t_{\rm ref}$ at Hanford and use the \texttt{LALSuite} library \cite{LAL} to compute the reference time in Livingston given the sky location. 

\begin{figure*}
\centering
\includegraphics[width=\linewidth]{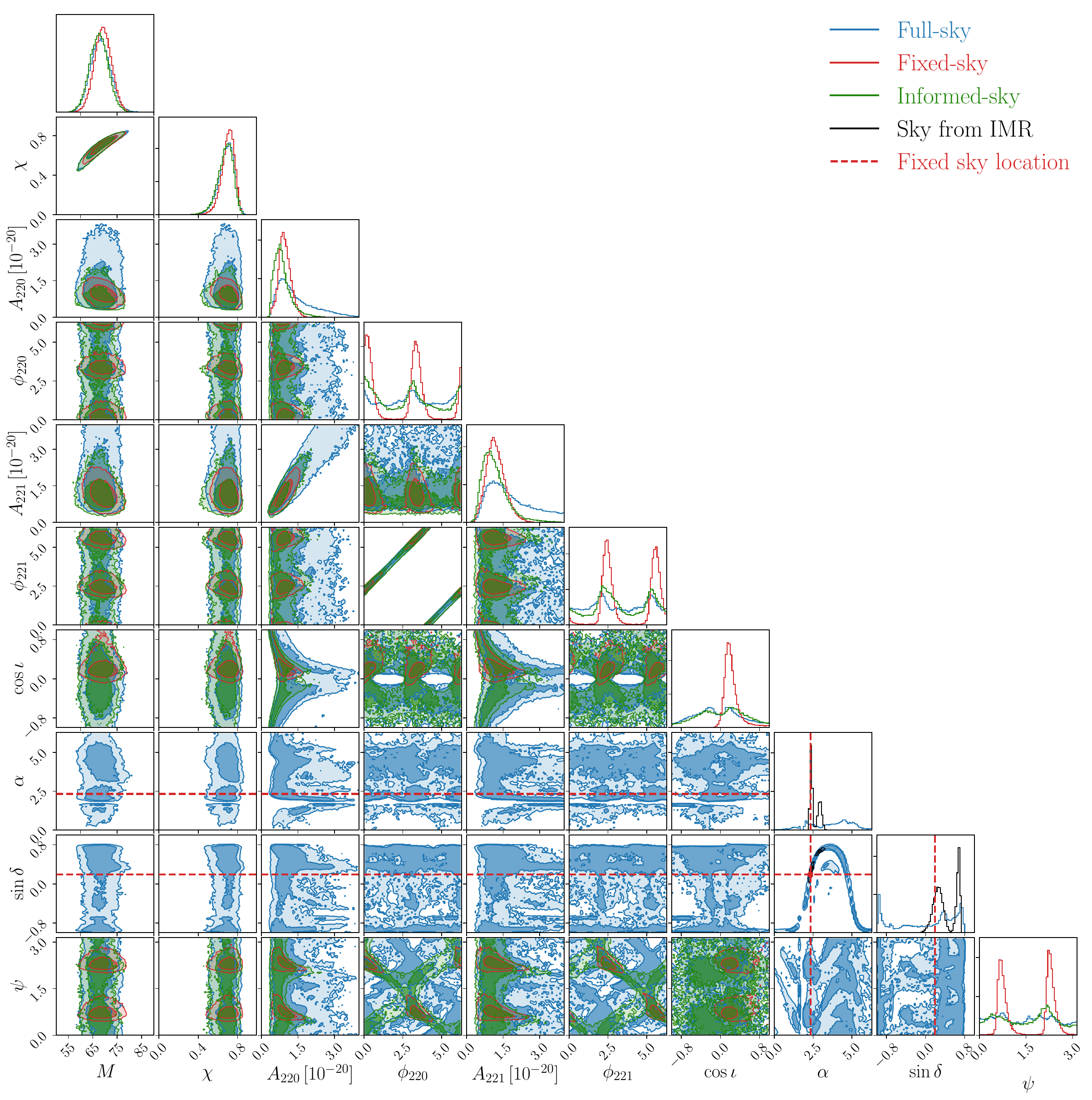}
\caption{Posteriors ($68 \%$ and $95 \%$ regions for 2D marginals) for GW250114 using a $(2,2,0)+(2,2,1)$ model, comparing the \FULL\ (blue), \FIX\ (red), and \HYB\ (green) analyses. The IMR posteriors for $(\alpha, \sin\delta)$ are in black and the fixed values used in \FIX\ are marked by a dashed red line.}
\label{fig:GW250114_comp}
\end{figure*}

Fig.~\ref{fig:GW250114_comp}
shows the resulting posteriors. For comparison, we also show the IMR sky posteriors $(\alpha, \sin\delta)$ used in \HYB. The \FULL\ posteriors are broader  except for mass and spin, and show the characteristic inverted \textit{U}-shaped $\alpha$--$\sin\delta$ correlation from inter-detector time delays.
\edits{Reassuringly, the amplitudes and phases inferred with the three approaches peak at similar values. However, the posteriors for \FIX\ are much narrower than in \FULL\ and even \HYB, as evident in the 1D marginal distributions and even more in the 2D contours. Moreover, \FIX\ also eliminates the multimodality in $\cos\iota$.
We compute the Bayes factor for the $(2,2,1)$ mode, finding $\ln \mathcal{B}^{221+220}_{220} \sim 7.42 \pm 0.02$ for \FULL, $ \sim 9.26 \pm 0.02$ for \FIX\ and $\sim 7.66 \pm 0.06$ for \HYB.} Thus, although \FIX\ overestimates the evidence for the overtone, the result remains robust to the treatment of extrinsic parameters, despite the loss of Livingston data in \FULL\ and \HYB.

Sampling the sky location is most useful when its IMR estimate is highly multimodal, e.g.\ for GW190521 due to its high mass. For this event, we model the signal with the $(2,2,0) + (3,3,0)$ modes.
We set the GPS reference time to $t_{\rm ref}=1242442967.4318974$, corresponding to $t_{\rm peak}+6$~ms $(\approx 4\ t_M)$ at Hanford (with $t_M = 325\ M_\odot = 1.6$ ms), i.e.\ where Ref.~\cite{Capano:2021etf} finds the highest evidence for the $(3,3,0)$ mode. For the \FIX\ analysis, we fix $(\alpha,\delta)=(3.5, 0.73)$ as in Ref.~\cite{Capano:2021etf}, although different studies have adopted different point estimates for $(\alpha, \delta)$ (see Table I of Ref.~\cite{Siegel:2023lxl}), and compute the corresponding starting time in Livingston. Here, we do not apply \HYB\ (see footnote \ref{foot}).

\edits{Notice that, while also Virgo detected GW190521, we do not include it in our analysis. This is because Virgo observes the signal $\sim 6.4\, t_M$ after the Hanford detector (using the IMR sky location). Since the \FULL\ prescription requires to cut the data at a common reference time set by the strain peak at the last-observing detector, including Virgo would sacrifice a significant portion of the data at the two LIGO sites.
Thus, although a third detector generally improves source localization, the significant loss in ringdown SNR does not justify its inclusion.}

\begin{figure*}
    \centering
    \includegraphics[width=\linewidth]{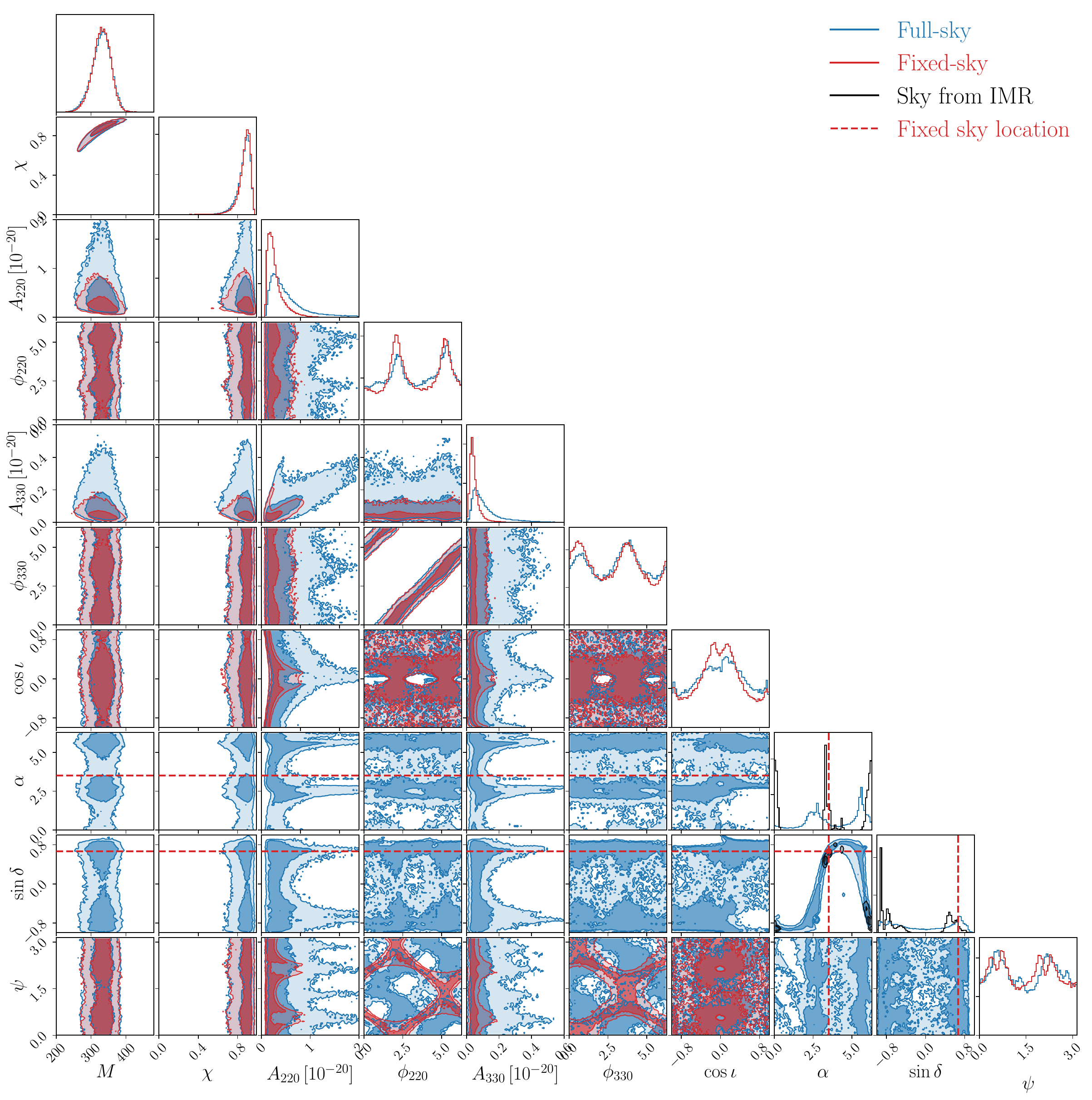}
    \caption{Same as Fig.~\ref{fig:GW250114_comp}, but for GW190521 using a $(2,2,0)+(3,3,0)$ model. \HYB\ is not applicable here (see text).}
    \label{fig:GW190521_comp}
\end{figure*}

GW190521 posteriors are shown in Fig.~\ref{fig:GW190521_comp}. \edits{This time, after marginalizing over the unknown sky location, \FULL\ amplitude and phase posteriors are only slightly wider than \FIX\ ones.} This is likely due to the similar and largely uninformative inclination and polarization posteriors in both analyses.
\edits{We also find that both approaches provide weak evidence for the higher mode, with $\ln \mathcal{B}^{330+220}_{220} \sim 2.44 \pm 0.02$ for \FULL\ and $\sim 0.82 \pm 0.02$ for \FIX}.
\edits{This illustrates that accounting for sky-location uncertainty does not necessarily reduce the Bayes factor for additional QNMs, as marginalization over the sky position can affect the competing waveform hypotheses differently.}

\begin{figure*}
\subfloat[$A_{221}/A_{220}$ posterior \label{fig:GW250114_amp_ratio}]{%
  \centering
  \includegraphics[height=5.2cm]{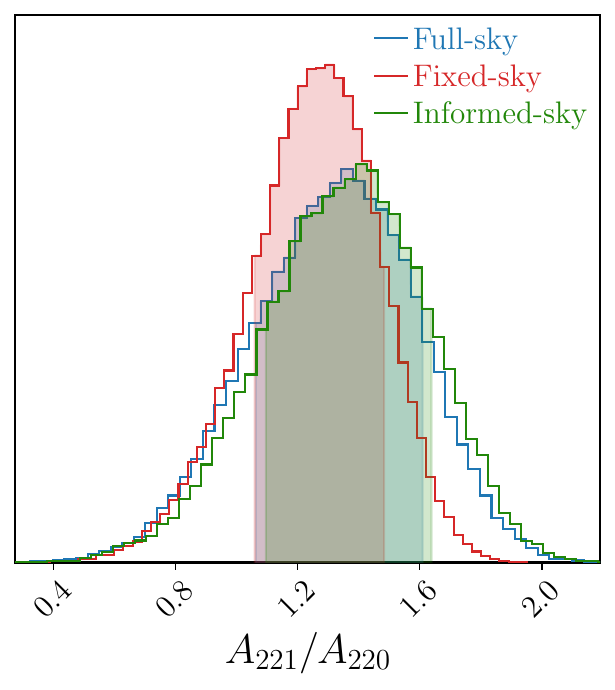}%
}\hfill
\subfloat[$A_{221}/A_{220}$ posterior and $\ln \mathcal{B}^{221+220}_{220}$ at different start times \label{fig:GW250114_violin}]{%
  \centering
  \includegraphics[height=5.2cm]{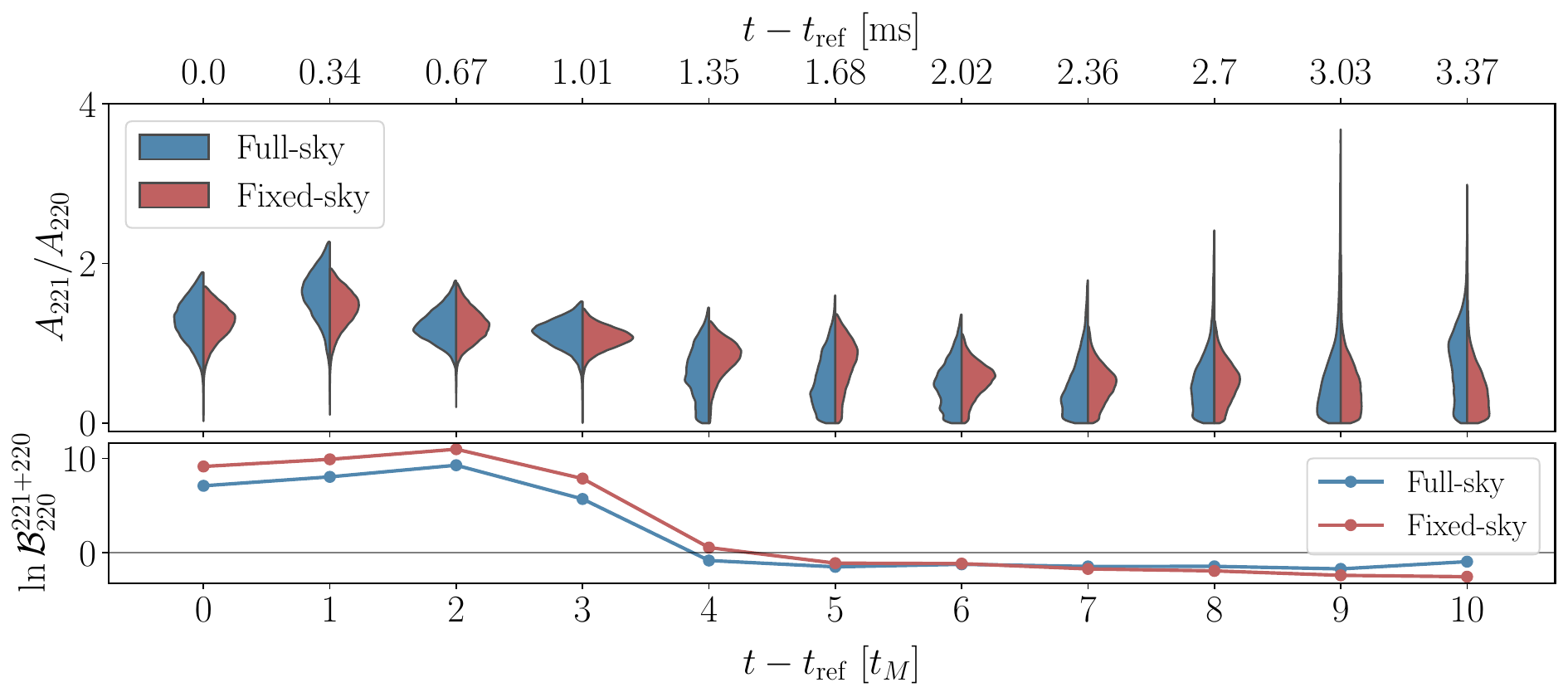}%
}
\caption{Posteriors for $A_{221}/A_{220}$ and $\ln \mathcal{B}^{221+220}_{220}$ in GW250114. (a) Comparison of the \FULL\ (blue), \FIX\ (red), and \HYB\ (green) approaches at $t = t_{\rm ref}$ ($68 \%$ shaded region). (b) \FULL\ (blue) and \FIX\ (red) posteriors and Bayes factors at different start times \edits{(relative to $t_{\rm ref}=t_{\rm peak}+6t_M$)}, computed at intervals of $t_M= 68.409 \ M_\odot = 0.337$ ms.}
\label{fig:GW250114_amp}
\end{figure*}

\begin{figure*}
\subfloat[$A_{330}/A_{220}$ posterior \label{fig:GW190521_amp_ratio}]{%
  \centering
  \includegraphics[height=5.2cm]{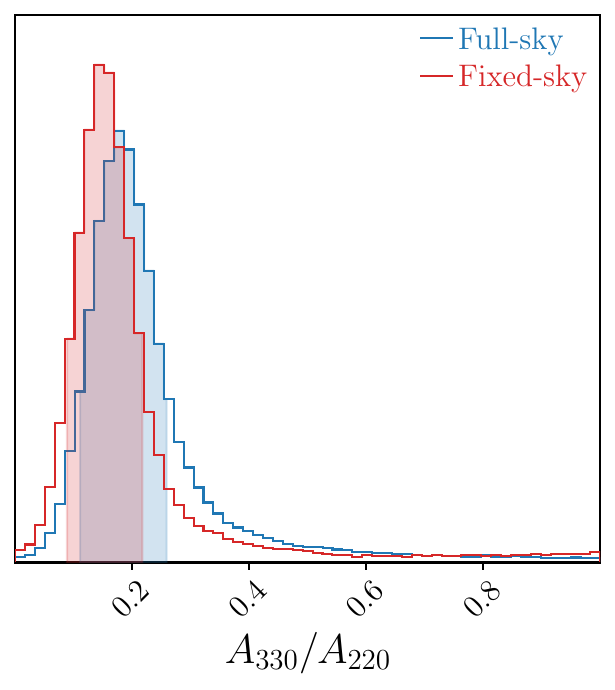}%
}\hfill
\subfloat[$A_{330}/A_{220}$ posterior and $\ln \mathcal{B}^{330+220}_{220}$ at different start times \label{fig:GW190521_violin}]{%
  \centering
  \includegraphics[height=5.2cm]{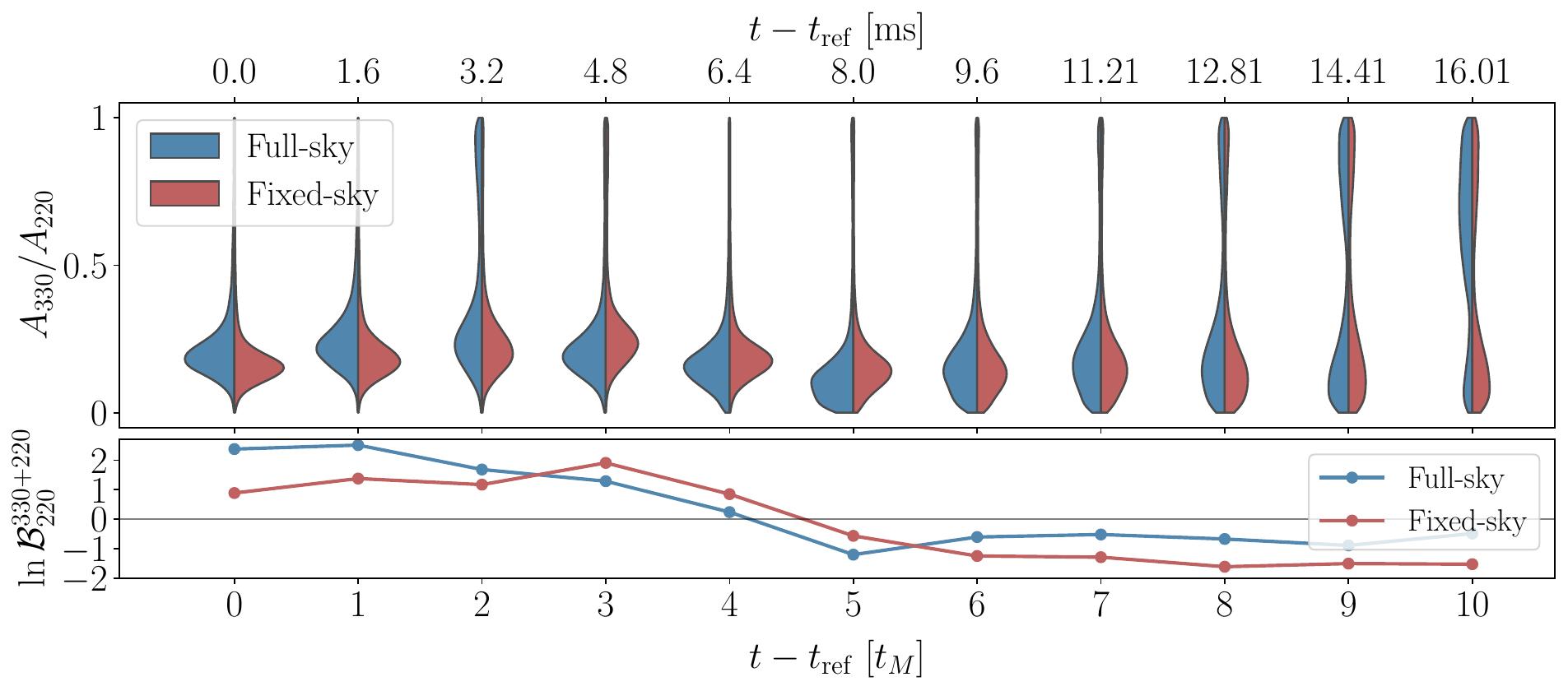}%
}
\caption{Posteriors for $A_{330}/A_{220}$ and $\ln \mathcal{B}^{330+220}_{220}$ in GW190521. (a) Comparison of the \FULL\ (blue) and \FIX\ (red) approaches at $t = t_{\rm ref}$ ($68 \%$ shaded region). (b) \FULL\ (blue) and \FIX\ (red) posteriors and Bayes factors at different start times \edits{(relative to $t_{\rm ref}=t_{\rm peak}+6$~ms)}, computed at intervals of $t_M = 325\ M_\odot = 1.6$ ms.}
\label{fig:GW190521_amp}
\end{figure*}

Figures~\ref{fig:GW250114_amp_ratio}--\ref{fig:GW190521_amp_ratio} show amplitude-ratio posteriors: all approaches yield similar distributions.
This suggests that, in \FIX\ analyses, amplitude ratios provide a more robust observable than individual mode amplitudes.

Figures~\ref{fig:GW250114_violin}--\ref{fig:GW190521_violin} show the temporal evolution of the amplitude ratios and Bayes factors. For GW250114, the ratio decreases and flattens, since the overtone decays faster than $(2,2,0)$, with $\ln \mathcal{B}$ becoming negative in both analyses about $10\ t_M$ after the peak, in agreement with Ref.~\cite{LIGOScientific:2025wao} (cf.\ their Fig.~10; \edits{note that $t_{\rm ref}+4t_M=t_{\rm peak}+10t_M$}). For GW190521 the $(2,2,0)$ and $(3,3,0)$ damping times are comparable and large (because of the high mass), so the ratio evolution is gradual, with support for zero only far from $t_{\rm ref}$; by contrast, $\ln \mathcal{B}$ becomes negative much earlier, indicating that the data already cease to support the presence of the additional mode even while the posterior remains broad and assigns little support to zero.

\edits{Finally, to quantify the extent to which \FIX\ (or \HYB) underestimates uncertainties relative to \FULL, we compute the ratio of posterior volumes in the space of the relevant intrinsic parameters, namely $\{M,\chi,A_{\rm 220}, A_{\rm 221 (330)}\}$.
The effective posterior volume is given by $V = e^H$, where $H$ is the differential entropy that we estimate using the Kozachenko–Leonenko k-nearest-neighbor estimator~\cite{Kraskov_2004} with $k=5$.
We find that $V_{\FIX}/V_{\FULL}=0.16$ and $0.12$ for GW250114 and GW190521, respectively, and $V_{\HYB}/V_{\FULL}=0.32$ for GW250114.
The fact that \HYB\ (which has the same data content as \FULL) reduces the volume by a factor of $\sim 3$ demonstrates that at least part of the reduction in $V_{\FIX}/V_{\FULL}$ is due to the sky uncertainty that \FIX\ neglects, rather than entirely to the amount of data lost in Livingston.}

\section{Conclusions}\label{sec:conclusions}

We have shown that fixing the sky location to point estimates in ringdown analyses produces posteriors on the QNM amplitudes (and phases) that are artificially narrow, as they do not account for sky-location uncertainty. By contrast, the \FULL\ treatment yields broader posterior distributions by properly marginalizing over the sky position.
Moreover, the latter avoids double-counting the sky-localization information already contained in the ringdown signal. 
\edits{When applicable, \HYB\ correctly accounts for the prior knowledge of the sky position (up to a less severe double-counting of the sky information)}, and therefore yields narrower posteriors than \FULL, while still avoiding the potential bias introduced by fixing the sky location to point estimates.
On the contrary, amplitude ratios are  consistent across approaches, making spectroscopic tests based on  relative amplitudes~\cite{Forteza:2022tgq} less sensitive to  extrinsic parameters.
For GW250114, the overtone detection is robust to the treatment of extrinsic parameters, although \FIX\ slightly overestimates the associated evidence. For GW190521, 
the Bayes factors \edits{are small for both approaches}, only mildly hinting at the presence of the $(3,3,0)$ mode.

Methodologically, we provide a statistically correct ringdown analysis, which samples {\it all} extrinsic parameters jointly with intrinsic ones; our JAX-based nested-sampling pipeline is remarkably efficient, completing a \FULL\ analysis with \edits{$10^4$ live points} in $\sim35$~minutes on an Apple M4 MacBook Pro.

\edits{We conclude by advocating for the use of our \FULL\ (or, alternatively, \HYB) method for unbiased and reliable QNM inference in current GW detections. This will be even more important for next-generation detectors~\cite{Yi:2024elj,Capuano:2025kkl}.}

\textit{Acknowledgments.} The code and accompanying tutorial notebooks are available at Ref.~\cite{repo_name}. The data necessary to reproduce all the figures in the paper are available on Zenodo at Ref.~\cite{dey_2026_20089762}.
We thank Gregorio Carullo and Joachim Pomper for useful discussions and comments on an earlier version of the draft.
KD acknowledges support from the NASA LISA foundation Science Grant 80NSSC19K0320. EB acknowledges support from the European Union's Horizon ERC Synergy Grant ``Making Sense of the Unexpected in the Gravitational-Wave Sky'' (Grant No.\ GWSky-101167314). MC is funded by the European Union under the Horizon Europe's Marie Sk{\l}odowska-Curie project~101065440.
This material is based upon work supported by NSF's LIGO Laboratory which is a major facility fully funded by the National Science Foundation. This research has made use of data or software obtained from the Gravitational Wave Open Science Center (\url{gwosc.org}), a service of LIGO Laboratory, the LIGO Scientific Collaboration, the Virgo Collaboration, and KAGRA.

\appendix

\section{Simulations}\label{app:sims}

\edits{To validate our pipeline, we perform a set of noiseless pure-ringdown and full IMR injections, so that any differences among the methods can be attributed solely to the treatment of the sky-location parameters rather than to noise artifacts. We start with ringdown-only injections using two representative cases, corresponding to GW250114- and GW190521-like events.} Specifically, we inject $(2,2,0)+(2,2,1)$ and $(2,2,0)+(3,3,0)$ modes, respectively.
The source parameters are:
\bea
&& \{M, \chi, A_{220}, \phi_{220}, A_{221}, \phi_{221}, \cos\iota, \alpha, \sin\delta, \psi\} = \nn \\
&& \{72\ M_\odot, 0.75, 0.5\cdot 10^{-20}, 2.2, 0.8 \cdot 10^{-20}, 4.5, 0.80, \nn \\
&& 2.33, 0.19, 1.33\} \,; \nn 
\eea
and
\bea
&& \{M, \chi, A_{220}, \phi_{220}, A_{330}, \phi_{330}, \cos\iota, \alpha, \sin\delta, \psi\}= \nn \\
&& \{300\ M_\odot, 0.8, 0.7\cdot 10^{-20}, 4.1, 0.17\cdot 10^{-20}, 1.5, 0.27, \nn \\
&& 3.5, 0.67, 1.33\} \,, \nn
\eea
with network optimal signal-to-noise ratios (SNRs) of $28.5$ and $25.1$, respectively. The common reference time adopted in \FULL\ reduces the SNRs to $23.3$ and $23.5$, respectively.

Here, we can only compare the \FULL\ and \FIX\ approaches, given the absence of a full IMR inference. In the \FIX\ analysis, we fix the sky position to the injected ground-truth values.
We have verified that fixing the sky position to different random values strongly biases all the other parameters except the mass and spin, whereas fixing it within the IMR sky posterior of the corresponding GW-like event does not lead to substantial changes in the posterior distributions.

\begin{figure*}
    \centering
    \includegraphics[width=\linewidth]{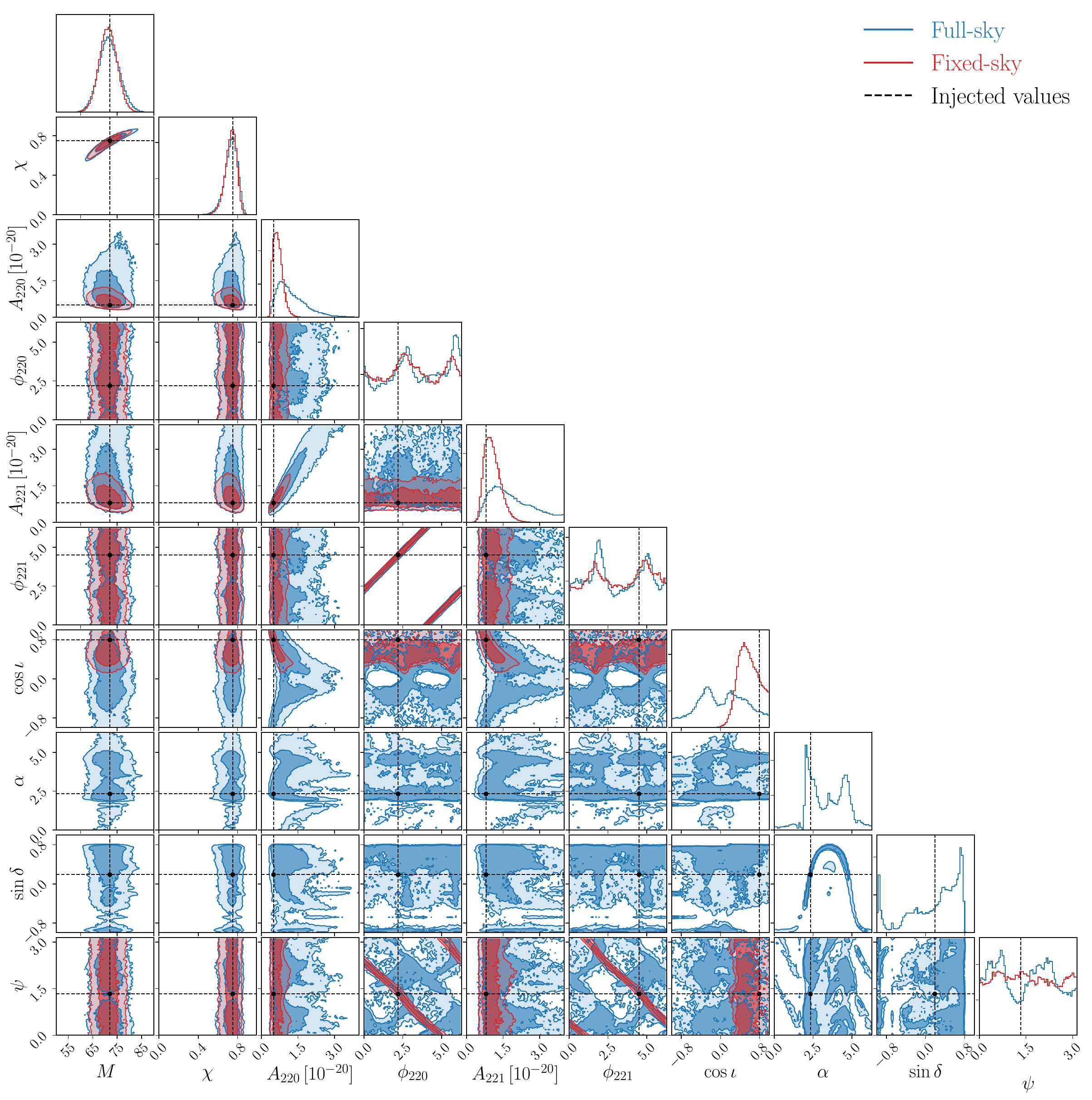}
    \caption{Posterior distributions ($68 \%$ and $95 \%$ regions for 2D marginals) for a GW250114-like simulated noiseless signal using a $(2,2,0)+(2,2,1)$ ringdown model, comparing the \FULL\ (blue) and \FIX\ (red) analyses. \FIX\ uses the injected values as sky position.}
    \label{fig:220-221-sim}
\end{figure*}

\begin{figure*}
    \centering
    \includegraphics[width=\linewidth]{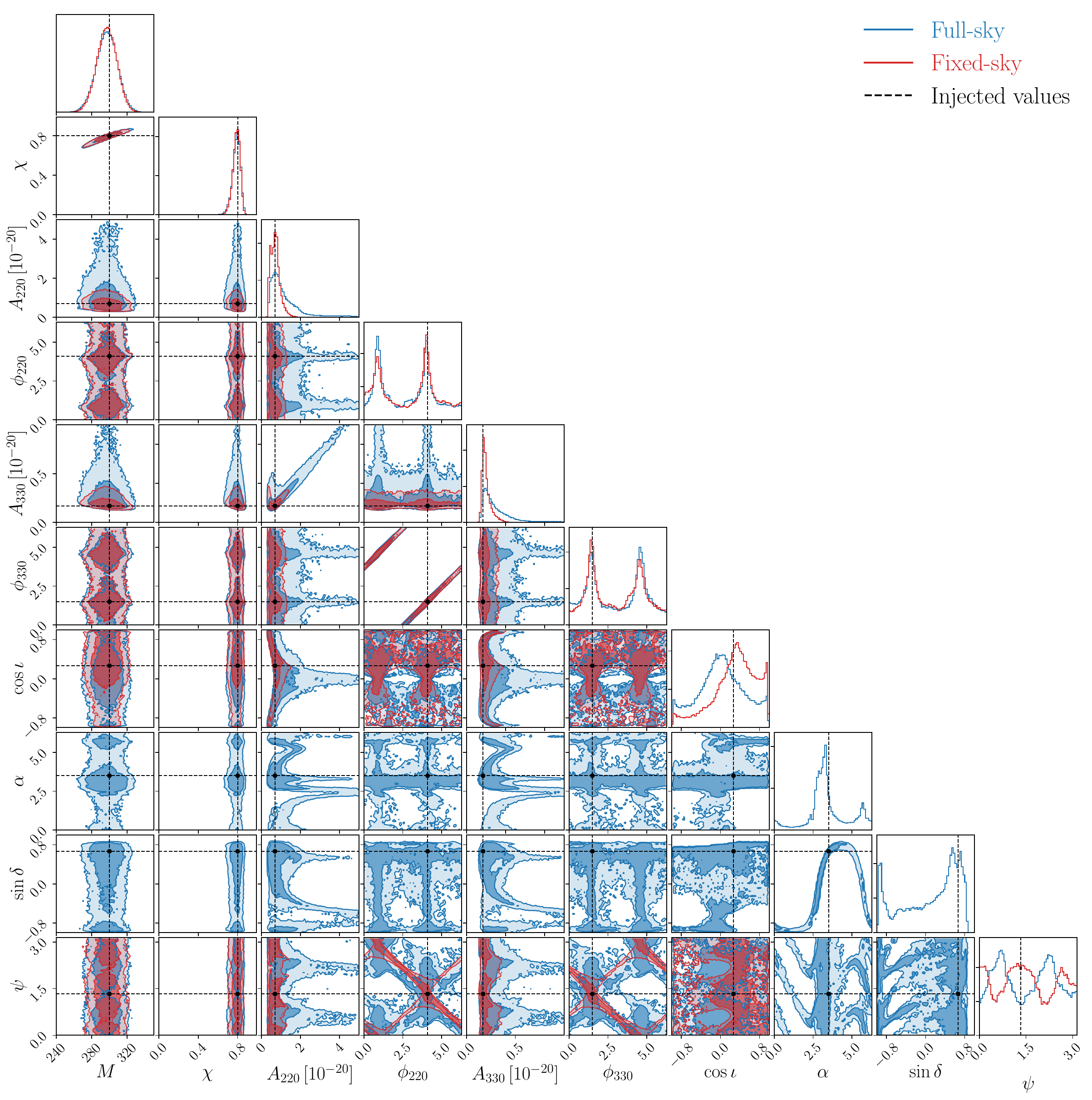}
    \caption{
    Posterior distributions ($68 \%$ and $95 \%$ regions for 2D marginals) for a GW190521-like simulated noiseless signal using a $(2,2,0)+(3,3,0)$ ringdown model, comparing the \FULL\ (blue) and \FIX\ (red) analyses. \FIX\ uses the injected values as sky position.
    }
    \label{fig:220-330-sim}
\end{figure*}

The posterior distributions are shown in Figs.~\ref{fig:220-221-sim} and \ref{fig:220-330-sim}, respectively.
We observe that mass and spin estimates are largely unaffected by the treatment of the sky location, whereas QNM amplitude posteriors become artificially narrow when the sky position is fixed.
We also show the posteriors of the amplitude ratios in Fig.~\ref{fig:noiseless-sim-amps}. For the GW250114-like event (left), the mean is unchanged but the variance is larger in the \FULL\ approach, likely due to the loss of SNR in Livingston. 
\edits{For the GW190521-like event (middle), by contrast, the combined effect of the loss of Livingston data on the one hand, and the absence of degeneracy breaking between amplitudes and extrinsic parameters due to different $(\ell,m)$ mode ratio on the other (see footnote~\ref{foot1}), results in good agreement between the two approaches.}

\begin{figure*}
    \centering
    \includegraphics[width=\linewidth]{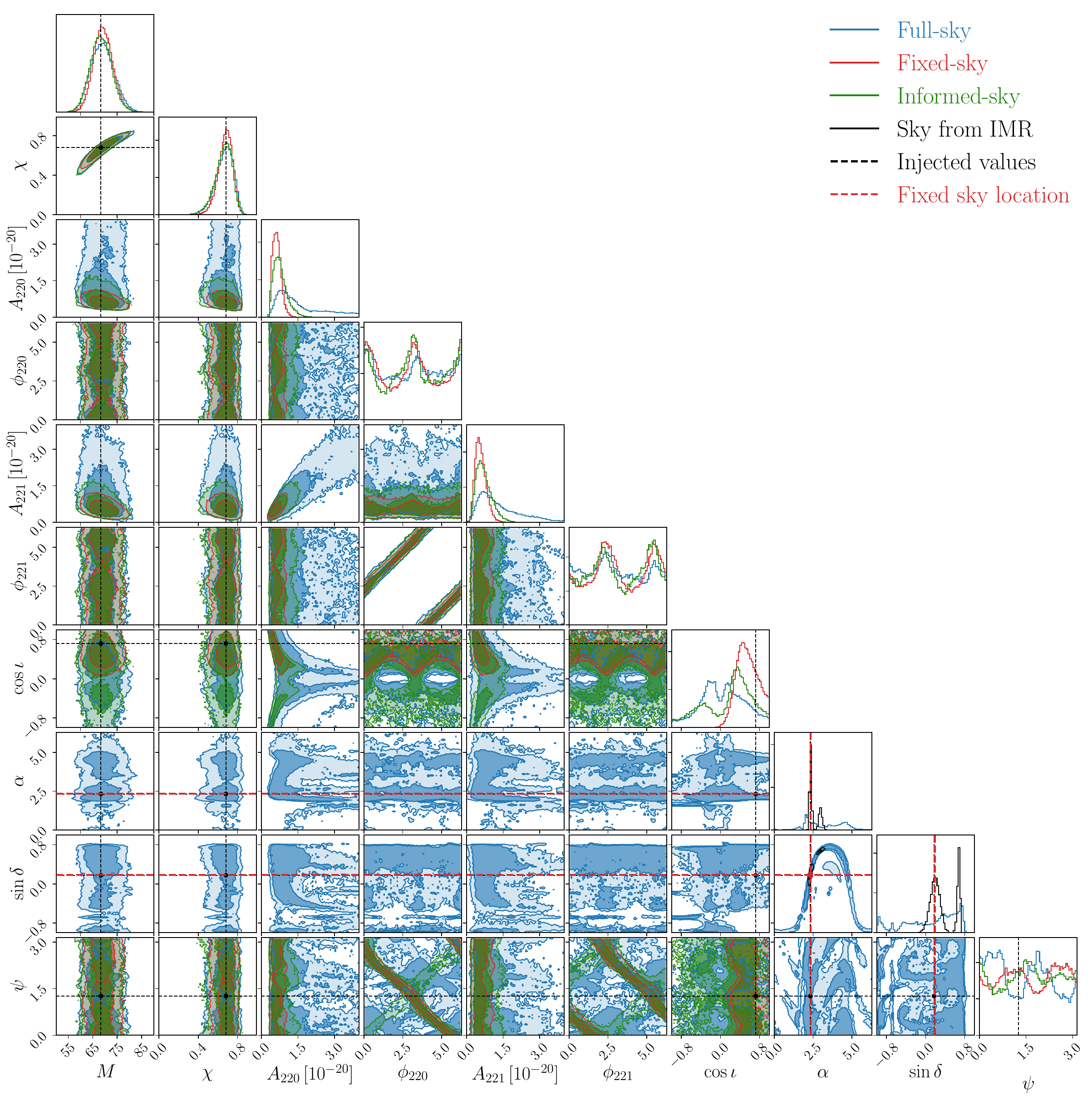}
    \caption{Posterior distributions ($68 \%$ and $95 \%$ regions for 2D marginals) for a GW250114-like NRSur7dq4 noiseless injection, using a $(2,2,0)+(2,2,1)$ ringdown model, comparing the \FULL\ (blue), \FIX\ (red) and \HYB\ (green) analyses. \FIX\ uses the IMR MAP values as point estimates for the sky position.}
    \label{fig:220-221-NRSur}
\end{figure*}

\begin{figure*}
\centering
\subfloat[\label{fig:220-221-sim-amp-ratio}]{%
  \includegraphics[height=5.2cm]{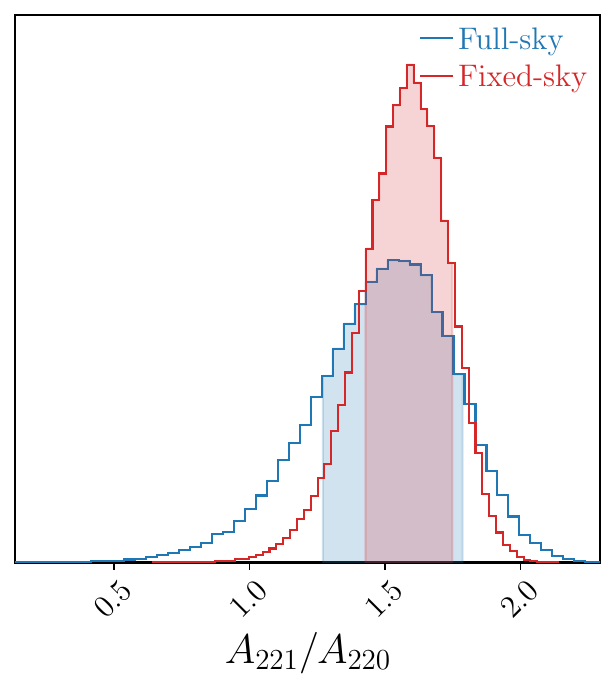}%
} \qquad\qquad
\subfloat[\label{fig:220-330-sim-amp-ratio}]{%
  \includegraphics[height=5.2cm]{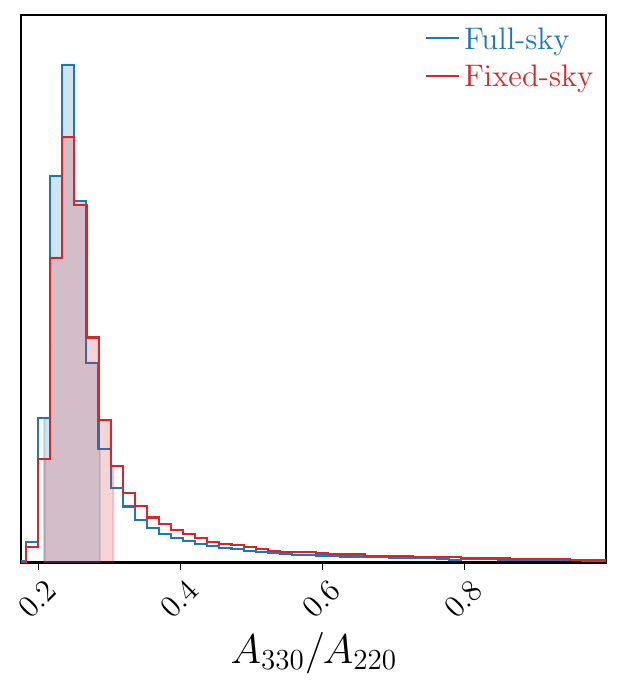}%
}\qquad\qquad
\subfloat[\label{fig:220-221-NRSur-amp-ratio}]{%
  \includegraphics[height=5.2cm]{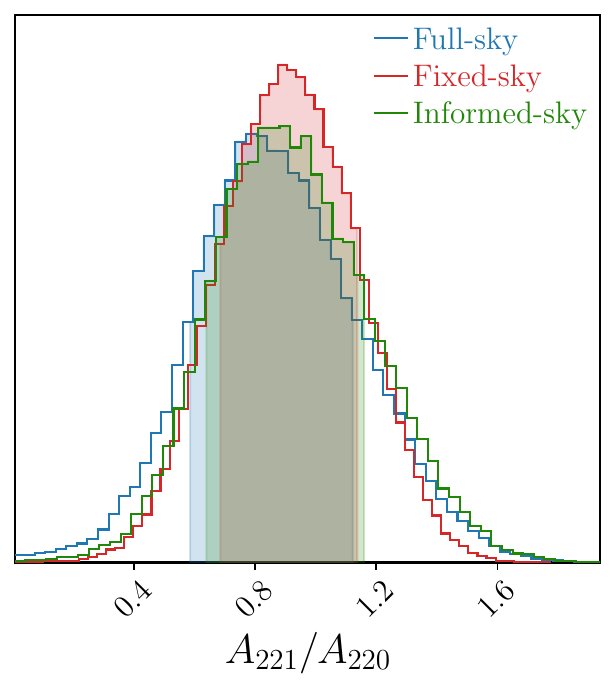}%
}
\caption{Posteriors ($68 \%$ shaded region) of the mode-amplitude ratios for the noiseless injections: (a) pure-ringdown injection of a GW250114-like event with $(2,2,0)+(2,2,1)$ modes, showing $A_{221}/A_{220}$; (b) pure-ringdown injection of a GW190521-like event with $(2,2,0)+(3,3,0)$ modes, showing $A_{330}/A_{220}$; (c) NRSur7dq4 injection of a GW250114-like event, showing $A_{221}/A_{220}$.}
\label{fig:noiseless-sim-amps}
\end{figure*}

\edits{To extend the validation to more realistic scenarios, we tested our methods on NRSur7dq4 injections generated using \texttt{LALSimulation}. Here, we present the results for a GW250114-like signal with the following parameters: 
\bea
&& \{m_1, m_2, \chi_{1z}, \chi_{2z}, d_L, \theta_{JN}, \phi_{\rm ref}, \alpha, \delta, \psi\} = \nn \\
&& \{36.2\ M_\odot, 35.6\ M_\odot, -0.01, -0.03, 392\ {\rm Mpc}, 0.76, \nn \\
&& 1.52, 2.33, 0.17, 1.26 \} \,. \nn
\eea
}
\edits{
The network SNR is 78. The parameter estimation is performed with \texttt{Bilby} and the resulting posteriors are used for the ringdown analysis. More precisely, the maximum a posteriori (MAP) for the geocentric merger time is used to set the starting time; the sky location posteriors are used in \HYB\ and their MAP estimate is used for the \FIX\ analysis. Fig.~\ref{fig:220-221-NRSur} shows the posterior distribution of the ringdown parameters, and Fig.~\ref{fig:220-221-NRSur-amp-ratio} shows the posteriors of the ratio $A_{221}/A_{220}$. The posterior volume is much larger for \FULL\ than for \FIX\ and \HYB\ ($V_{\rm \FIX}/V_{\rm \FULL}\sim 0.05,\, V_{\rm \HYB}/V_{\rm \FULL}\sim 0.17)$. Moreover, since the signal is noiseless, the amplitude ratios are centered on the same value, with \FIX\ having narrower variance than \FULL\ and \HYB, likely due to the data loss in the latter approaches.
}

\bibliographystyle{./utphys}
\bibliography{main}

\end{document}